\documentstyle[aasms4,11pt]{article}

\slugcomment {Centennial Issue of {\it The Astrophysical Journal}, in press}

\begin{document}
\title
{Commentary on "The Theory of the Fluctuations in Brightness of the Milky
Way. V" by S. Chandrasekhar and G. Munch (1952)}

\author
      {John Scalo}
\affil {University of Texas at Austin}
\authoremail{parrot@astro.as.utexas.edu}

\begin{abstract}
	The series of papers by Chandrasekhar and Munch in the 1950s were
concerned with the use of statistical models to infer the properties of
interstellar clouds based on the observed spatial brightness fluctuations
of the Milky Way.  The present paper summarizes the subsequent influence
of this work, concentrating on the departure from their earlier discrete
cloud model to a continuous stochastic model in Paper V of the series.
The contrast between the two models anticipated and parallels current
tensions in the interpretation of interstellar structure, as well as
intergalactic Lyman alpha clouds.  The case of interstellar structure is
discussed in some detail.  Implications concerning the reification of
models and the ability of scientific abstraction to model complex
phenomena are also briefly discussed.
\end{abstract}

\pagebreak


	To present-day astrophysicists studying the spatial structure of
the interstellar medium, the
intergalactic medium, or the large-scale distribution of galaxies in the
universe, a methodology
involving the comparison of the spatial statistics of a model with the
observed statistics seems
standard.  But in fact this approach to interstellar structure, and the
models employed, trace back to
the work of Ambartzumium in the 1940s (see Ambartzumian 1950 and Kaplan \&
Pikelner 1970, p.173), and
were developed most extensively in a series of papers by Chandrasekhar and
Munch in the 1950s
(hereinafter CM, Papers I through V).  These papers and their results were
influential in a number of
subsequent works.  For example, the relation between the two-point spatial
correlation function of
galaxies and the angular correlation function, which played a dominant role
in the study of large-scale
structure in the 1970s and 1980s (before the availability of large redshift
surveys and the recognition
of the need for additional structure descriptors), involves an equation
derived by Limber (1953a, see
Peebles 1993, pp.216-217) and inspired by the formulation of CM's  1952
Paper V.  For the interstellar
medium, several studies of interstellar reddening and HI emission and
absorption have been used, assuming
a discrete cloud model, to infer the mean number of clouds per unit length
along a line of sight and the
mean extinction per cloud, based on results given in the CM papers (e.g.
Knude 1979).  Perhaps the most
significant aspect of this early work, however, is the manner in which the
rather radical departure from
their earlier discrete cloud model to a continuous stochastic model in
Paper V parallels current tensions
in the interpretation of both the interstellar medium and even
intergalactic Lyman alpha clouds, as
explained below.

	Prior to their 1952 paper, CM presented four papers that were
concerned with inferring a few basic
properties of a discrete cloud model from the observed spatial fluctuations
in the brightness of the
Milky Way.  The brightness fluctuations were interpreted as being due
primarily to the varying number of
discrete absorbing clouds along a line of sight.  These papers are
mathematically daunting, but the basic
ideas are simple and have direct relevance today.  In Paper I CM derived an
integro-differential equation
for the brightness fluctuations in terms of the frequency distribution of
cloud extinctions.  This
equation can be seen as a specific example of the Chapman-Kolmogorov
equation describing the probability
distribution of variables that undergo both continuous changes as well as
"jumps."  (A simpler
approximate derivation can be found in Kaplan \& Pikelner 1970.)  In
subsequent papers, they solved this
equation for the cases in which the system of clouds has infinite extent
and a particular distribution of
extinction (Paper II), the case of finite extent but constant extinction
per cloud (Paper III), and the
general case of arbitrary distributions of extinction for infinite extent
(Paper IV).   Limber (1953b)
generalized to finite extent, while Munch (1955, Paper VI) used the model
to estimate the decorrelation
length in the Milky Way.  All of this work assumed a model in which clouds
are discrete entities.

	A substantial departure, however, is seen in the fifth, 1952, paper
V, in which CM considered replacing
the discrete  cloud model by a continuous  stochastic model for the density
field.  This continuous model
was generalized to finite extent by Ramakrishnan (1954).  As CM say in
their abstract, the new picture
"Šmay be considered as an alternative to (or a refinement of) the current
picture, which visualizes the
interstellar medium as consisting of a distribution of discrete clouds."
Besides presaging the study of
structure in terms of spatial statistics defined for a continuous random
variable, this paper resonates
with current work suggesting that the density distribution of Lyman alpha
absorbers at intermediate and
large redshift (Bi \& Davidson 1997, Croft et al. 1998, Haehnelt et al.
1998) and of cool interstellar
gas (e.g. Falgarone 1990 for an observational review, Ballesteros-Paredes
et al. 1999 for a theoretical
discussion) should be viewed as a continuous, albeit intermittent, density
and velocity field rather than
as a collection of discrete entities called "clouds."  The utility of the
discrete cloud model in
furnishing quantitative results is not in question.  The question is
whether the discrete cloud model
omits some basic physics that is essential to understanding the evolution
of the gas and its ability to
form stars and galaxies, or leads to an account of such processes that is
simply incorrect at some
(possibly severe) level.  A useful analogy is to consider geologists trying
to understand mountain
formation and evolution by studying only the peaks high enough to receive
snow at some given time, as
though they were separate entities.  Could such an approach ever discover
the fundamental role of
stress-driven "wrinkling" of the Earth's crust in the evolution of mountain
ranges?

The assumption by CM of discrete clouds in Papers I-IV reflected the
general belief among astronomers at
the time that the apparent discreteness in {\it velocity} space of the
absorption lines seen toward OB
stars, catalogued most comprehensively by Adams (1949), reflected a
discrete {\it spatial} structure.
Since that time such a correspondence has become "clouded" by evidence that
the spectral lines are usually
blends of multiple components, and the realization that these velocity
features only represent a small
fraction of the density structure that later became accessible to observers
using a number of other
techniques.  The observed morphology of interstellar gas and dust is now
acknowledged to be much more
complex than allowed for by the discrete cloud model, in terms of the
geometry of density enhancements,
nested hierarchical structure, and connectedness.  Other authors have
argued against the discrete cloud
model on grounds other than morphology.  For example, Dickey \& Lockman
(1990), in their review paper on
neutral hydrogen in our Galaxy, discuss "Šthe difficulty of objectively
delimiting discrete features in
emission surveys" in connection with the discrete cloud model.  Falgarone
et al. (1992), in a study of
small-scale molecular cloud structure, use the modest inferred density
contrasts to conclude that "...the
overall pictureŠresembles that of the interstellar medium described in 1952
by Chandrasekhar and Munch as
a continuous distribution of density fluctuations of small amplitude at
each scale...rather than an
ensemble of discrete clouds."  There is also substantial evidence that the
density and velocity fields
are scale-free over a large range of scales (see Elmegreen \& Falgarone
1996 and references therein),
suggesting a continuous distribution (although a system of discrete clouds
with a continuous distribution
of sizes or tiny discrete clouds of the same size arranged hierarchically
could also explain the
observational results).  Nevertheless, the effect of sensitivity selection
effects and the usually poor
extent of spatial sampling has continued to reinforce the prevalent
interpretation of observations in
terms of a collection of discrete, (usually) spherical, or at least smooth,
clouds.  Any contour map that
is based on less than of order ten thousand spatial resolution elements,
and which has a sensitivity
limit that omits a significant areal fraction of the region being studied,
will appear to be a system of
more-or-less spherical discrete clouds.  What has changed is only an
increase in the number of categories
of discrete clouds, such as diffuse, dark, giant molecular, clumps, cores,
etc., which usually reflect
the distinct observational techniques and resolutions employed more than
clear evidence of real category
boundaries.

There is another, just as significant, fact that has helped entrench the
discrete cloud model:  the model
is intrinsically easier to visualize and to model theoretically.  For
example, evolution of structure in
the discrete model can be abstracted into a generalized "coalescence
equation," in which the clouds
interact only through collisions, reducing the interstellar medium to a
more complex version of the
kinetic theory of gases.  This model was first treated in some detail by
Field \& Saslaw (1965), and by
a large number of subsequent works (an especially illuminating treatment
can be found in Silk \&
Takahashi 1979).  There is little connection to the flows on scales larger
than an individual cloud size,
or coupling in velocity space, except through collisions.  Concerning the
origin of the clouds, an even
simpler model pictures discrete clouds that continually condense by thermal
instability, resulting in a
conceptually static two-phase model for the interstellar medium (Field,
Goldsmith, \& Habing 1969).
Even though this model has been generalized to include a hot phase
involving supernova explosions, as
well as other effects, most notably in the influential paper by McKee \&
Ostriker (1977), the
discreteness assumption for the clouds, and the weak connection to the
hydrodynamics, persists.

In contrast, if the discreteness assumption is abandoned, one is left,
theoretically, with having to
infer statistical properties from the hydrodynamic equations for a
turbulent magnetized gas, a problem
that defies satisfactory solution, let alone conceptualization, even in
studies of laboratory and
terrestrial unmagnetized incompressible turbulence.  The problem lies in
the complex and unpredictable
behavior of the nonlinear advection operator in the momentum equation.  The
problem of empirical
description is also made more difficult, since instead of counting objects
with various attributes, less
intuitive statistical descriptors must be employed.  Even conceptually, the
connection to the fluid
equations makes the picture inherently more dynamic and difficult to
visualize.  (However the continuous
CM 1952 model did not address the dynamics, since they did not consider the
available information [Adams
1949] concerning the velocity field.  The interpretation of the dynamics in
terms of the stochastic model
was apparently first treated by Kaplan and by von Hoerner in the 1950s--see
Kaplan \& Pikelner 1970.)
The tension between the complexity of empirical and simulated turbulence,
on the one hand, and the many
attempts to reduce it to a conceptual model, can be seen as a major theme
in the history of the study of
turbulence.  It seems very likely that the continuous stochastic variable
alternative proposed in CM's
1952 paper was influenced by Chandrasekhar's active involvement in
turbulence theory around the same
time.

Therefore, while providing the basic conceptual basis for the discrete
cloud model  for several
subsequent generations of astronomers studying reddening, extinction, HI,
molecular line, and more
recently submillimeter continuum observations, this series of papers culminates by asking whether their
own discrete model should be replaced by a tu
rbulence-like formulation of the problem of interstellar
structure, a question which can be seen as a major theme in contemporary
studies of the interstellar
medium.  Although CM (1952) apparently favored the continuous stochastic
approach, they gave virtually no
guidance concerning criteria by which this approach could be evaluated
relative to the discrete cloud
model, except for brief reference to the degree to which the models can
match observations (see also
Limber 1953b).  In fact it is only relatively recently that attention has
returned (from the time of
Kaplan and von Hoerner in the 1950s) to descriptions in terms of continuous
density and velocity fields,
using, for example, the correlation function (see Miesch \& Bally 1994 and
references therein),
functions related to wavelet transforms (Langer et al. 1995), and principal
component analysis (Heyer \&
Schloerb 1997).  For the most part, however, theoreticians and observers
alike treat the interstellar gas
as some version of the discrete model, which has its theoretical and
visualistic advantages, as outlined
above.

Part of the continued attraction of the discrete model is its conceptual
simplicity, which is usually
seen as a positive value in much astrophysical research and science in
general.  Yet CM were presciently
aware of the negative reification effects which accompany the discrete
cloud model:  "We wish to
emphasizeŠa tendency to argue in circles can be noted in the literature in
that confirmation for the {\it
picture}  [italics in original] of interstellar matter as occurring in the
form of discrete clouds is
sought in the data analyzed"  (CS 1952, p. 104).  Such reification
continues to the present day, despite
occasional warnings.

The fact that CM did not make an explicit decision between the two models
based on either empirical or
theoretical considerations can be interpreted as acknowledgement of the
utility of {\it some form} of an
abstraction, or reduction, of the irregular, dynamic, complex, magnetized,
and undeniably continuous,
interstellar turbulent flow.  Indeed, short of simulations, which cannot be
regarded as explanatory
theories {\it per se}, the nature of science seems to require some  kind of
abstraction to a conceptual
model in order to generalize from the particular to the universal.  The
interesting questions that
therefore arise from the series of papers by CM are: 1. What kind of
conceptual model can best bridge the
gap and how can we avoid reifying the model?  2. Is {\it any} kind of
conceptual model capable of
bridging the gap?  These questions address issues usually relegated to the
philosophy or sociology of
science.  But it seems clear that they should be of concern to scientists
in general, since the first
question calls for a new approach to the problem at hand, while the second
question challenges the basis
of the traditional scientific enterprise in the sense that it questions the
idea that complex physical
phenomena can be adequately described by universal models.  Whether CM
recognized these broader
implications in their shift from the discrete to the continuous model is of
course unknown.  Yet the fact
that their two models still in effect drive most of the contemporary work
in the field of interstellar
(and intergalactic) structure and star formation, and that the above
questions have not yet been answered
in the nearly five decades since their work appeared, underscores the
fundamental importance of the CM
papers, and provides continued motivation for the many astronomers who are
trying to bridge the gap in
order to understand the evolution of gas and star formation in galaxies.



\end{document}